\documentclass[runningheads]{llncs}
\usepackage[T1]{fontenc}
\usepackage{graphicx}
\usepackage{lipsum}
\usepackage{subcaption}
\usepackage{caption}
\usepackage{multirow}
\usepackage{makecell}
\usepackage{hyperref}
\usepackage{svg}
\usepackage{booktabs}
\usepackage[square,sort,comma,numbers]{natbib}
\usepackage{placeins} 
\usepackage{url} 

\begin{document}
\title{A Dataset for Automatic Vocal Mode Classification}
%

\author{Reemt Hinrichs\inst{1}\orcidID{0000-0003-3352-7223} \and
Sonja Stephan\inst{1} \and
Alexander Lange\inst{1}\orcidID{0009-0005-0737-371X} \and
Jörn Ostermann\inst{1}
\authorrunning{R. Hinrichs et al.}
%
\institute{$^1$ Institut für Informationsverarbeitung, Leibniz University Hannover, Hannover, Germany
\email{hinrichs@tnt.uni-hannover.de}\\
}}
\maketitle              
\begin{abstract}
The Complete Vocal Technique (CVT) is a school of singing developed in the past decades by Cathrin Sadolin et al.. CVT groups the use of the voice into so called vocal modes, namely Neutral, Curbing, Overdrive and Edge. 
Knowledge of the desired vocal mode can be helpful for singing students.
Automatic classification of vocal modes can thus be important for technology-assisted singing teaching.
Previously, automatic classification of vocal modes has been attempted without major success, potentially due to a lack of data.
Therefore, we recorded a novel vocal mode dataset consisting of sustained vowels recorded from four singers, three of which professional singers with more than five years of CVT-experience. The dataset covers the entire vocal range of the subjects, totaling 3,752 unique samples. By using four microphones, thereby offering a natural data augmentation, the dataset consists of more than 13,000 samples combined.
An annotation was created using three CVT-experienced annotators, each providing an individual annotation.
The merged annotation as well as the three individual annotations come with the published dataset.
Additionally, we provide some baseline classification results.
The best balanced accuracy across a 5-fold cross validation of 81.3\,\% was achieved with a ResNet18. The dataset can be downloaded under \url{https://zenodo.org/records/14276415}.

\keywords{complete vocal technique \and vocal modes \and  artificial intelligence \and convolutional neural network \and singing voice dataset }
\end{abstract}
\section{Introduction}
\label{introduction}
Singing voice analysis and more specifically vocal technique analysis is commonly employed by singing teachers as well as singing students, either as a first step in assisting a student to improve their voice (or performance) or to assess how to achieve a certain sound \cite{Miller96}.
Due to limited time and budget, time spent with a singing teacher is usually a scarce resource. Additionally, singing teachers are attention constrained, i.e. likely to miss or forget certain details when giving their feedback to a student.
These two points could be improved on through automatic singing voice analysis.
Automatic singing voice analysis aims to algorithmically replicate some of the analyses performed by singing teachers, either for entertainment purposes (usually simple pitch estimation) or to assist singing students and teachers in their work. 
Thanks to the explosion of deep learning in the past decade, the field has seen some growth in the recent years. A good introduction is given by Humphrey et al.  \cite{Humphrey19}.
Terminology in singing pedagogy can be confusing  \cite{Bernardoni08, McGlashan17}. A common example being support or belting, which can be difficult to grasp for early voice students.
Different vocal pedagogy "schools" exist, especially in the context of contemporary commercial music, which provide different ways to think and speak about singing and come with different ideas on how to approach the development of ones singing voice. Examples are Estill Voice Training \cite{Fantini17},  Speech Level Singing \cite{McClellan11} but also the Complete Vocal Technique (CVT) developed by Cathrin Sadolin \cite{CVTbook}. 
CVT is a system that attempts to describe  the entire range of sounds the human voice can healthily produce as well as providing rules and advice on how to actually achieve these sounds.
Vocal modes, introduced by CVT and which will be described in Section \ref{ssec:cvt}, are a useful way of thinking about certain characteristics of the voice. These vocal modes  ought to refine for instance the vocal category belting \cite{McGlashan17}.

\textbf{Aim and Contribution}
While some advances have been made in automatic singing voice analysis, the field is still impeded by a lack of publically available datasets compared to, e.g., computer vision. 
A well known dataset for singing voice analysis is VocalSet \cite{VocalSet} which consists of arpeggios sang in different styles by professional singers.
However, the voice traits recorded for the VocalSet cannot (or hardly) be used to analyze the singing voice with respect to CVT terminology. 
For the example of CVT, which provides useful categories of certain qualities of the singing voice, no dedicated datasets have been published to date.
Automatic classification of vocal modes has been previously attempted \cite{Brixen14, Sol23} but so far with unsatisfying results and without publishing a dedicated dataset. This considerably hinders the application of machine learning algorithms.
Because the automatic classification of vocal modes can be a helpful tool for singers \cite{Sol23}, in this work we present a novel dataset. The dataset is made up of samples recorded from four singers covering their entire vocal range for all four vocal modes defined by CVT, namely Neutral, Curbing, Overdrive and Edge. 
Besides aiding research on vocal mode classification, the dataset can be helpful to similar singing voice analysis tasks such as distinguishing belting from non-belting.

\section{Method and Materials}

\subsection{The Complete Vocal Technique}
\label{ssec:cvt}
The Complete Vocal Technique (CVT) is a school of voice pedagogy developed by Cathrin Sadoline  \cite{CVTbook}  that aims to provide research-backed descriptions and instructions on how to  create any sound of the human voice found in music healthily \cite{McGlashan23, CVTWebsite}. Some of its elements are found under different names in other voice pedagogies.
Its three basic principles are support, "necessary twang", and avoiding protrusion of the jaw \cite{CVTbook}. 

\textbf{Vocal Modes}
CVT categorizes the human voice into four distinct classes - the vocal modes - namely Neutral, Curbing, Overdrive, and Edge. The vocal modes are determined by the properties or qualities "hold", twang, and "metal".
These three qualities can in principle be quantified, e.g., one can say that there is more or less metal/hold/twang in a voice, albeit usually a rather coarse quantification is used.
The term "hold" corresponds to the bodily feeling of holding back the sound, resulting in a restrained sounding voice, which can be heard, for example, in moaning. "Metal" refers to a sharper, brighter trait of certain sounds typically heard in  shout-like sounds. "Twang"
is a term commonly used in different schools of voice pedagogy and refers to a duck-like sound color \cite{Aaen21}.
Neutral is the vocal mode without any metal. The two other properties can be present without altering the vocal mode. 
Curbing is a reduced metallic mode, i.e., the voice contains some metal and a certain amount of hold. It does not use twang.
Overdrive is a full metallic mode in contrast to the reduced metallic mode Curbing. Edge is a full metallic mode which, in contrast to Overdrive, adds twang to yield a very sharp, bright sound often found in, e.g., hard rock belting \cite{Brixen12}.
\begin{table}[t]
    \centering
        \caption{List of vowels for each vocal mode according to the Complete Vocal Technique (CVT). The chosen list of vowels followed the official CVT book \cite{CVTbook}. Note that the most recent CVT book (and the corresponding smartphone app) slightly altered the Curbing vowels.}

    \resizebox{\linewidth}{!}{
    \begin{tabular}{|c|c|c|c|}
        \hline
         \makecell{Curbing} &  \makecell{Edge} & \makecell{Neutral} &  \makecell{Overdrive} \\
        \hline
\makecell{'I' (as in 'sit'), \\ 'O'  (as in 'woman'), \\'UH' (as in 'hungry')} & \makecell{'I' (as in 'sit'), \\ 'EH' (as in 'bed'), \\ 'A' (as in 'cat'), \\'OE' (as in 'herb')}  & \makecell{ \\'EE' (as in 'see' [i]), \\ 'OO' (as in 'food' [u]),\\ 'AH' (as in 'father'),\\ 'I' (as in 'sit'),\\ 'EH' (as in 'bed' ${\epsilon}$),\\ 'A' (as in 'cat' [ ae]),\\ 'O' (as in 'woman'),\\ 'OH' (as in 'so' [o])\\ \\} &\makecell{'EH' (as in 'bed'),\\ 'OH' (as in 'so')} \\
        \hline
    \end{tabular}}
    \label{tab:vowels}
    \vspace{-5mm}
\end{table}
However, because of the so called necessary twang, which is a certain amount of generally required minimal twang  to produce a sound, and  CVT's idea of "density", which aims to describe  more or less full sounding vocal modes (e.g. a reduced sounding/reduced-density Overdrive) \cite{Leppavuori21}, the distinctions between the vocal modes are gradual, similar to a color ray, which has regions of uncertainty between clearly distinct colors. Consequently, the distinction between the vocal modes becomes increasingly unclear as one deviates from the so called "center" of the mode. 
Roughly speaking, the "center" of a mode is the archetypal way of singing/producing a vocal mode. 
Except for Overdrive, which for males is limited from above by C5 and for females by D5, CVT does not impose limits (regarding healthy production) on the vocal range for the vocal modes. Specifically, there is no lower bound for the vocal range for any of the vocal modes.
CVT imposes vowel restrictions for the different vocal modes, that is, except for Neutral, in general only a certain subset of all possible vowels can be used for a vocal mode. These subsets are summarized in Table \ref{tab:vowels}. 
\subsection{Dataset}
\label{ssec:dataset}
The only research known to the authors attempting to automatically classify the four vocal modes of CVT was performed by the Complete Vocal Institute (CVI) \cite{Brixen13, Sol23}. However, no dataset was published and as such  no dedicated dataset for vocal mode classification exists. Therefore a novel data set was recorded in this work.
\begin{table*}[b]
    \centering
        \caption{Comparison of the nominal and the corresponding annotated vocal mode. For Curbing and Overdrive, a larger number of samples was annotated as Neutral.}
        \begin{tabular}{c|c|c|c|c|}
            Nominal \textbackslash Annotated & Neutral & Curbing & Overdrive & Edge\\
            \hline
            Neutral & 1580 & 20 & 4 & 1\\
            Curbing & 101 & 471 & 20 & 29\\
            Overdrive & 104 & 22 & 462 & 71\\
            Edge & 13 & 31 & 71 & 723\\
        \end{tabular}
        \label{tab:confusion_matrix_annotation}
\end{table*}
For this, four vocalists, labeled s1, s2, s3, and s4, were recorded at the Institut für Informationsverarbeitung, two males and two females. 
The subjects s1, s3 and s4  have at least 15 years of singing experience each, s2 has about five years of experience.
The age range covered 37 to about 48 years. 
All subjects gave their informed consent to publication of the recorded data.  The initial plan was to record six subjects. However, it proved to be impossible to find two additional subjects. Except for s2, who is the main author of this manuscript, all were professional singers with university level training and experience as vocal coaches or musical singers. s2 was a hobby singer experienced in CVT through his vocal lessons. One Behringer B-5 condenser microphone, one Stage Line ECM-40 condenser microphone, one IPhone, and one MotorolaOne smartphone were used for recording. The first two were used to record at high quality with slightly different frequency responses, the latter two to capture the audio in a more realistic scenario.
The smartphones' purpose was to provide a natural augmentation of the recordings, which can be crucial to achieve generalization for machine learning algorithms. 
The basic setup was meant to approximate the approach described in \cite{Brixen13}.
The condenser microphone recordings were sampled at 48 kHz, but were later downsampled to 44.1 kHz to match the IPhones sampling rate. The MotorolaOne recordings, recorded natively at 16 kHz, were also resampled to 44.1 kHz.
Sampling bit depth was 16 bits in all cases. The condenser microphones were recorded using a Focusrite 18i20 audiointerface and Cubase 9.5. 
Each subject sang major arpeggios across their entire vocal range, starting at the highest note they felt comfortable with in the respective vocal mode.
The major arpeggios were gradually reduced by one half step such that the entire vocal range of the singers was captured without gaps. This way, five such iterations per vowel were performed, each starting a semitone lower.
CVT imposes restrictions on the allowed vowels for each vocal mode as shown in Tab. \ref{tab:vowels}. Only in Neutral all vowels can be used. The number of allowed vowels is considerably reduced in the other modes. 
To maintain at least an approximately similar number of samples per mode, each vowel was repeated twice for Curbing and Edge, and three times for Overdrive. 
To steer the pronunciation of the vowels, each subject was provided with a list of vowels per vocal mode together with a (German) key word containing the respective vowel. 
Before singing the respective arpeggios, the subjects were presented corresponding major chords played by the experiment conductor, s2.
During annotation, a few recorded samples were found to have improper vocal oscillation. These samples were removed/excluded from the main dataset, but are provided in the online repository to assist possible future research on, e.g., singing mistakes.
After data polishing, the final annotated dataset is made up of 1605 unique nominal productions\footnote{In this work, the term "production" is used to refer to the original, unique vowels sang/produced by the subjects. Because several microphones were used to capture each production, the individual productions correspond to more than one sample.} for Neutral, 642 unique nominal productions for Curbing, 664 unique nominal productions for Overdrive and 841 unique nominal productions for Edge. As such, the entire dataset is made up of 3752 unique productions. 
As mentioned, most of these productions were recorded using in total four microphones, resulting in a total of 13335 samples.
The nominal vocal modes are the vocal modes the subjects were asked to produce. The nominal mode is not necessarily identical with the annotated mode. For example, the production of a vocal mode can fail, and the resulting recording/sound might be perceived by a listener/annotator as belonging to a different mode than intended. For example, near the lower end of a singer's range, maintaining the desired vocal mode can be challenging (with the exception of Neutral). 
While previous works such as \cite{Sol23} do not make this distinction, it is crucial for developing classification algorithms, which are supposed to map audio samples to the appropriate vocal mode. Such a classification has to be based on the actual sound of the samples (as perceived by a listener).
Each sample of the dataset consists of a single note, in almost all cases without onset and offset. The recorded audio data was manually cut for the Behringer recordings, and the well-known locally normalized cross-correlation method \cite{Yoo09} was used to automatically extract corresponding samples for the three other data sources/microphones. While this automatic extraction was not manually checked for correctness in its entirety, performed isolated comparisons between the automatically and manually extracted samples showed a 100\,\% agreement. Furthermore, a vast number of automatically extracted samples were listened to without finding a single obviously incorrectly extracted sample. 
The dataset, including the annotation, is available at \url{https://zenodo.org/records/14276415}, published under a Creative Commons Attribution Non-Commercial 4.0 International  license.\\
\textbf{Annotation}
\label{ssec:annotation}
Annotation was done by s1 (the vocal teacher of s2 and second author of this manuscript); s3 and an additionally invited annotator, because s4 declined the invitation to annotate the dataset. The invited third annotator was a voice student familiar with CVT through university course work as well as additional choir work.
Peak-normalized samples were passed to the annotators in randomized, consecutively numbered sequences, with each annotator receiving a different order. Thus the annotators operated on files labeled '0001.wav', '0002.wav' etc.
Annotation was performed without the presence of the main author during the private time of the respective annotators.
Random repetitions of previous samples, about 50 per annotator, were used to assess intra-rater reliability. 
The three individual annotations are included in the dataset to be hopefully combined with additional annotations in the future. 
For usage in classification tasks, however, a unique, merged/final annotation had to be constructed from the individual annotations. Because each annotator annotated/rated about 50 samples twice,  there are about 50 samples per individual annotation where the respective annotator gave two ratings, which were not necessarily identical. 
A single sample occurred for which two annotators each gave two ratings. This sample, nominal Neutral, was rated by all annotators and all ratings as Neutral.
Considering this, the three individual annotations received from the individual annotators were combined/merged in the following way: 
For each sample, all ratings of the annotators were aggregated, with the consequence, that up to four\footnote{Except for the mentioned sample with five ratings.} ratings had to be considered. The fourth rating was due to random repetitions as mentioned in the previous paragraph.
\begin{figure}[t]
    \centering    \includegraphics[width=\linewidth]{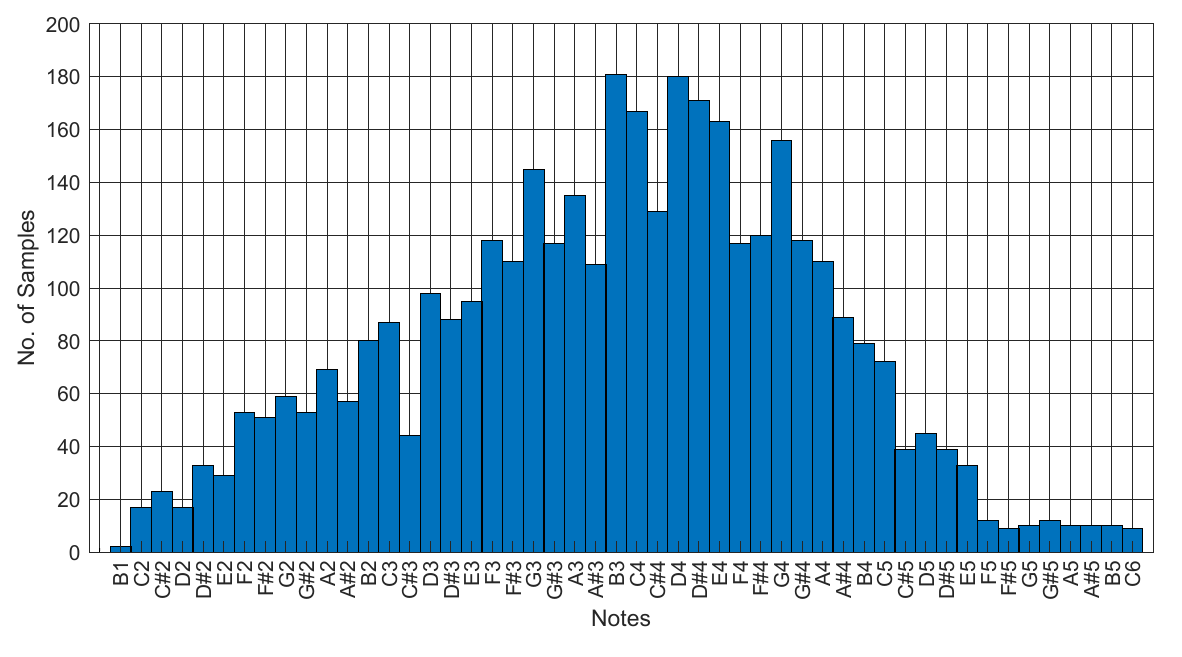}
    \vspace{-9mm}
    \caption{Number of samples per note for the entire dataset. The highest note, a C6, was achieved in nominal Neutral. The lowest note, a B1, was achieved also in nominal Neutral.  }
    \label{fig:total_notes}
\end{figure}
If the annotators reached a majority consensus for this aggregated annotation, that is, one of the four vocal modes was annotated by the majority (4:0, 3:1 or 2:1) of the aggregated annotations, this vocal mode became the annotated, ground truth vocal mode of the sample.
For 367 samples, no majority was found this way.
For these 367 samples, the nominal mode was used as a tie-breaker, but only if the nominal mode agreed with  a majority (2:2 or 1:1:1) of the annotators. This allowed to resolve all but 29 of these 367 samples.
These final 29 samples were deemed undecidable and marked accordingly in the final, merged annotation.
As such, these 29 samples were left out in the classification tasks reported in this work.
This merged annotation, consisting of 3723 samples, alongside the three individual annotations, is included in the published dataset.\\
\textbf{Annotation Metrics}
\label{sssec:annotation_metrics}
The raw agreement, i.\,e., without considering class imbalances, between the annotated and the nominal modes was 86.9\,\%. The balanced agreement between the annotated and the nominal modes, which considers class imbalances, was 82.7\,\%. The latter first computes the agreement separately for each class and then averages across the individual agreements.
These values are in agreement, both quantitatively and qualitatively, with previous results \cite{Brixen14, Sol23} - \cite{Sol23} reports a balanced agreement of 80.7\,\% - which showed that even professionals can disagree to some extent regarding the vocal mode of a vocal segment. 
This issue will be addressed in the discussion.
A correspondence table of the nominal and annotated vocal modes is given in Tab. \ref{tab:confusion_matrix_annotation}.
Intra-rater agreements were 0.68, 0.74 and 0.77. Inter-annotator agreements were 0.61, 0.63 and 0.65. 
The achieved Fleiss’ kappa score was 0.45.
Inter-annotator agreements as well as the Fleiss' kappa score were evaluated based on samples with exactly one rating by each involved annotator.\\
\textbf{Annotation Subsets}
A substantial difference in accuracy was observed, when training the classifiers using the previously described annotation as ground truth, compared to using the nominal modes as a natural, alternative ground truth. To get additional insight into the impact of the annotation on the resulting classification accuracy, the following two subsets of the annotation were considered:
A strong consensus subset, that only consisted of those samples rated identically by all annotators, i.e., all annotators assigned the same mode.
Similarly, a  weak consensus subset was constructed, where exactly one differing rating was allowed, that is, out of N ratings, at least N-1 were identical.
The number of unique samples was 1843 in the strong consensus subset and 3355 in the weak consensus subset (multiplied by the number of recording devices).
These subsets can be motivated by the previously mentioned idea of the center of a vocal mode as well as the observed disagreement of the annotators for certain samples. If, for example, a subject produced a mode, but partially deviated from its center, the distinction to another vocal mode can become less clear for this production. Especially in those cases, it is conceivable that the annotators gave contradicting ratings despite the acoustic features being quite similar.
In such a case, classification algorithms may be confused by contradicting annotations; that is, annotations which give different class assignments despite identical features.
If this was the case, a substantial improvement in classification accuracy is to be expected, at least for the strong consensus subset, seeing that one disagreeing rating can be seen as evidence for a deviation from a mode's center.
\begin{figure}[t]
    \centering
        \begin{subfigure}{0.49\textwidth}
\includegraphics[width=\linewidth]{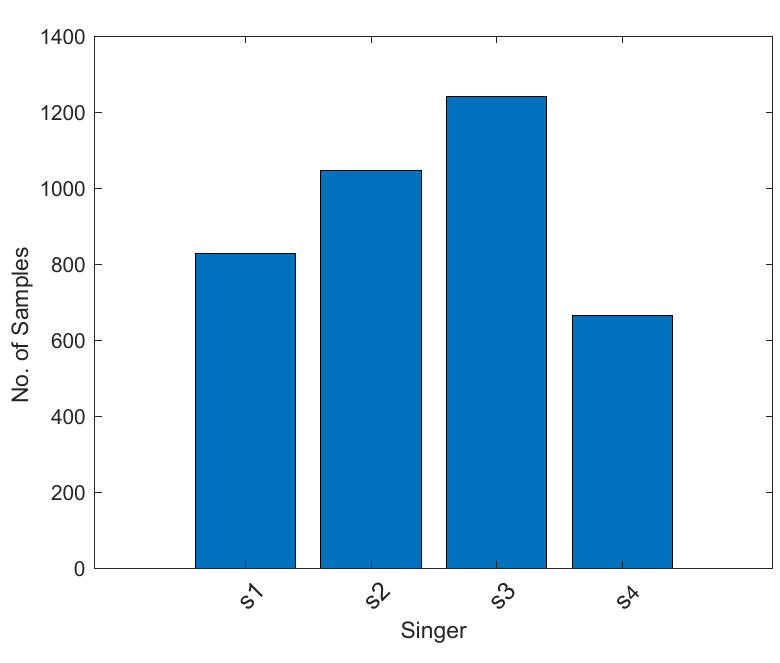}
    \caption{Number of samples per subject.}
    \label{fig:samples_per_singer}
    \end{subfigure}
    \begin{subfigure}{0.49\textwidth}
        \includegraphics[width=\linewidth]{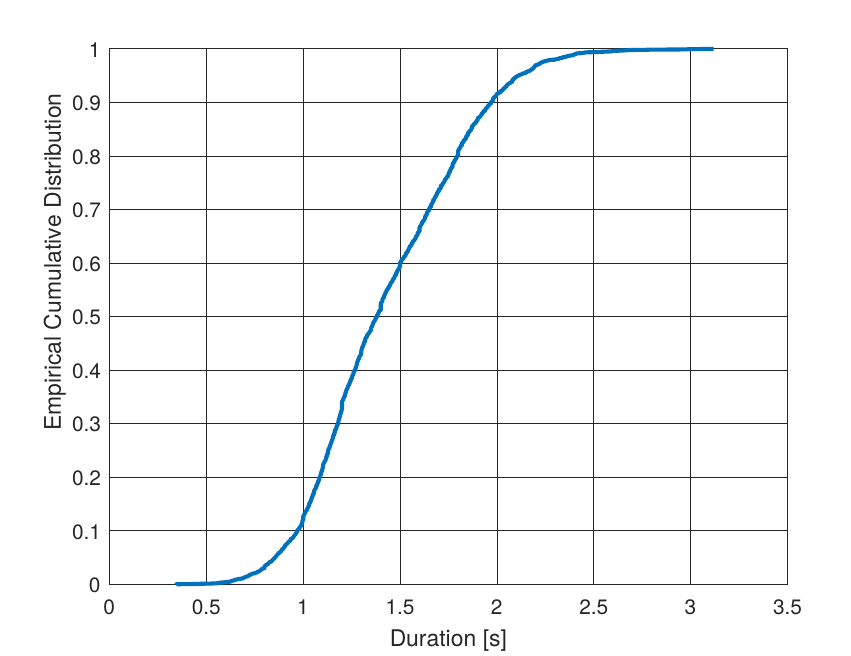}
    \caption{\small Distribution of the sample durations. }
    \label{fig:duration_dataset}
    \end{subfigure}
        \caption{Number of samples per subject and empirical cumulative distribution of the sample durations in seconds of all files of the dataset.}
        \label{fig:samples_per_singer_and_distribution}
\end{figure}
\\\textbf{Dataset statistics}
\label{ssec:statistics}
Fig. \ref{fig:total_notes} depicts the number of samples per note in the dataset. Above E5 only a minor number of samples were recorded, largely due to limitations in vocal range of the subjects (partially also due to upper limits for the vocal modes). The subject s3 had a very low voice, achieving an F1 as his lowest note, the lowest among all participants. This subject's highest note was an E5.
Similarly, the lowest and highest notes for subject s1 were a D\#3 and C6, for subject s2 an E2 and a D\#5, and for subject s4 an E3 and an E5. 
The number of samples per subject as well as the distribution of the sample duration in seconds is  given in Fig. \ref{fig:samples_per_singer_and_distribution}. The subject s3 had the largest vocal range, resulting in the largest number of samples. 
About 80\,\% of all recordings have a length between 1 and 2 seconds. 
\subsection{Baseline Classifiers}
To provide baseline classification results, several classification algorithms were trained and evaluated using the recorded dataset, namely  xgboost (XGB), a support vector machine (SVM), a k-nearest neighbor classifier (KNN), a random forest classifier (RF) and two  convolutional neural networks (CNNs), namely ResNet18 and ResNet34 \cite{ResNet}. 
Both finetuning the ResNets initialized with pretrained ImageNet weights and training newly, randomly initialized ResNets were tested. 
\begin{figure*}[t]
    \centering
    \begin{subfigure}{.49\textwidth}
\includegraphics[width=\linewidth]{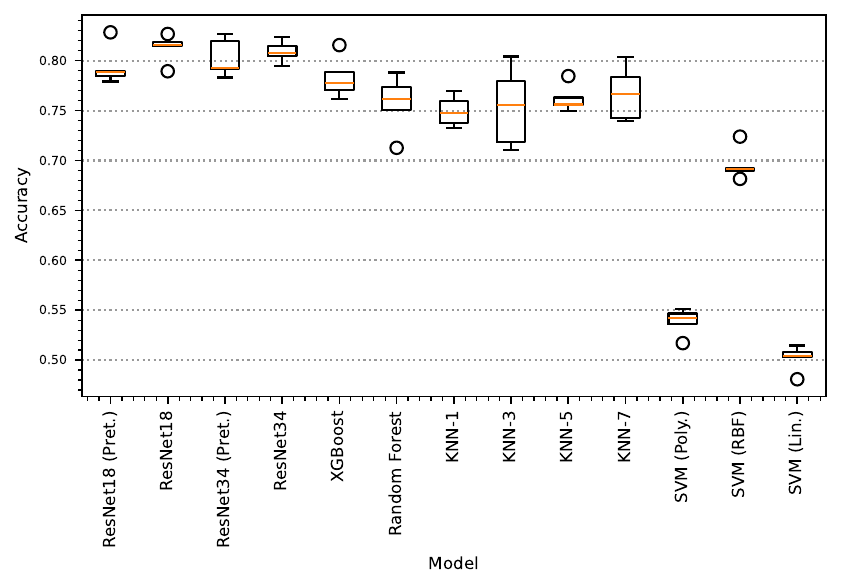}
\caption{Using Annotated Modes}
\label{fig:annotation_crossvalidation_boxplots_annotated}
\end{subfigure}
\begin{subfigure}{.49\textwidth}
\includegraphics[width=\linewidth]{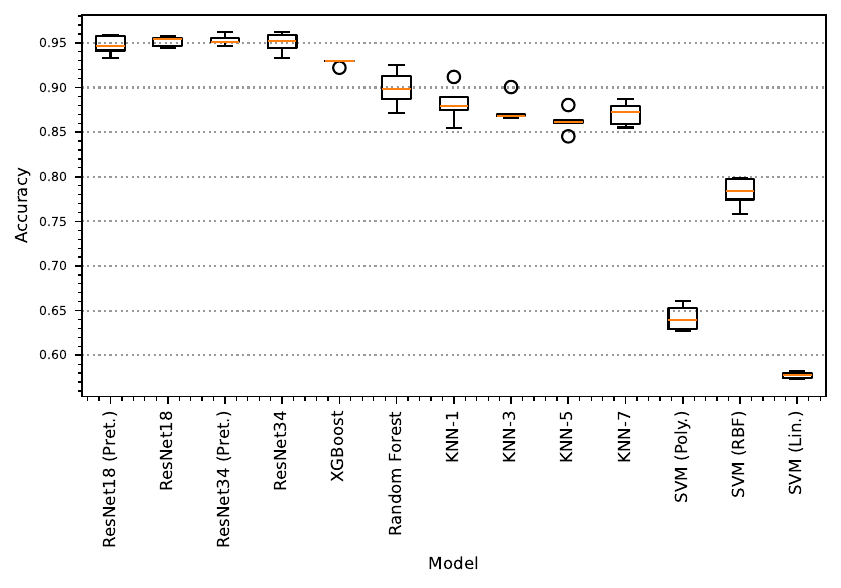}
\caption{Using Nominal Modes}
\label{fig:nominal_crossvalidation_boxplots_annotated}
\end{subfigure}
    \caption{Balanced accuracies on the test set across the 5-fold cross validation of all investigated classifiers using (a) the annotated vocal mode and (b) the nominal vocal mode as ground truth. }
\label{fig:crossvalidation_boxplots_annotated}
\end{figure*}
Non-CNN classifiers were implemented using the python library scikit-learn and default parameters, except for the KNN, where the number of considered neighbors was varied, as well as the random forest, which used a maximum tree depth of 30. 
XGB was implemented using the code published by \cite{Chen16}.
ResNets were implemented with pytorch 2.3.0 using default parameters if not stated otherwise.
\\\textbf{Data Preprocessing, Training and Evaluation}
In all cases, waveforms of the samples were loudness normalized and subsequently energy normalized\footnote{Only for training the baseline classifiers. The samples in the dataset were stored as is.}. 
In general, Overdrive and Edge exhibit greater signal energy than Neutral and Curbing. However, in a practical situation, the distance to the microphone can vary, and consequently the energy of the recorded signal. This energy difference should therefore not be used, implicitly or explicitly, as a feature for classification.
Additional energy normalization was motivated by observed slight differences in energy between the vocal modes even after loudness normalization. 
Initial learning rate was $5\cdot 10^{-5}$ and the learning rate was reduced by a factor of 0.8 every three epochs if the training loss had not improved. Stochastic gradient descent served as optimization algorithm. The batch size was set to 128. 
Due to the recording method, each vowel production corresponds to several audio samples (recorded with different microphones.
Therefore, splitting of the dataset into train and test subsets was performed such that no vowel production occurred in the training set that also occurred, recorded with a different microphone, in the test set. Splitting naively would distort the results, as the classifier would see augmented versions of training samples in the test set, i.e., the same production but different microphone.
For all classifiers, a 90\,\%/10\,\% train/test split was used; that is, 90\,\% of the dataset was used to train the classifier, whereas the other 10\,\% was used solely for evaluating the trained model.
The performance of the classifiers was measured using balanced accuracy across a 5-fold cross validation.

\vspace{-1mm}
\section{Results}
\label{sec:results}
Boxplots of the test balanced accuracies for all investigated classifiers across the 5-fold cross validation are depicted in Fig. \ref{fig:crossvalidation_boxplots_annotated}. 
Fig. \ref{fig:annotation_crossvalidation_boxplots_annotated} depicts results when using the annotation and Fig. \ref{fig:nominal_crossvalidation_boxplots_annotated} shows results when using the nominal modes as ground truth. Note the difference in y-axes' ranges. 
Corresponding test balanced accuracies, computed from the aggregated confusion matrices of the 5-fold cross validation, are given in Tab. \ref{tab:balanced_accuracies}.
\begin{table}[t]
    \centering
        \caption{Balanced accuracies for all classifiers across the 5-fold cross validation on the test set and for the phone subset of the test set. Shown are the results for all four conditions considered: using the full annotation ("Annotated"),  using the nominal modes as ground truth, using only the strong consensus subset and using the weak consensus subset.
        }
        \resizebox{\linewidth}{!}{
   \begin{tabular}{|l|cc|cc|cc|cc|}
\hline
\multirow{2}{*}{Classifier} 
  & \multicolumn{2}{c|}{Annotated} 
  & \multicolumn{2}{c|}{Nominal} 
  & \multicolumn{2}{c|}{Strong Consensus} 
  & \multicolumn{2}{c|}{Weak Consensus} \\
  & All [\%] & Phones [\%] 
  & All [\%] & Phones [\%] 
  & All [\%] & Phones [\%] 
  & All [\%] & Phones [\%] \\
\hline
ResNet18 (Pret.)     & 79.5 & 78.9 & 94.7 & 93.7 & 90.6 & 89.2 & 81.9 & 81.2 \\
ResNet18             & \textbf{81.3} & \textbf{80.9} & 95.2 & 94.2 & 89.4 & 86.5 & \textbf{82.4} & \textbf{81.4} \\
ResNet34 (Pret.)     & 80.3 & 79.9 & \textbf{95.3} & \textbf{94.4} & \textbf{90.9} & \textbf{89.8} & 81.5 & 80.6 \\
ResNet34             & 80.9 & 80.5 & 94.9 & 93.6 & 87.1 & 85.1 & 79.6 & 78.9 \\
XGBoost              & 78.2 & 77.0 & 92.8 & 91.5 & 84.4 & 82.6 & 78.7 & 77.9 \\
Random Forest        & 75.7 & 74.1 & 89.9 & 88.7 & 76.7 & 73.2 & 75.3 & 74.4 \\
KNN-1                & 74.8 & 74.6 & 88.2 & 88.0 & 80.0 & 78.9 & 74.6 & 74.4 \\
KNN-3                & 75.2 & 74.6 & 87.4 & 86.2 & 76.2 & 77.1 & 74.3 & 72.9 \\
KNN-5                & 76.1 & 75.8 & 86.2 & 85.2 & 79.2 & 78.8 & 75.7 & 75.1 \\
KNN-7                & 76.7 & 76.5 & 87.0 & 86.1 & 77.4 & 74.4 & 75.9 & 75.5 \\
SVM (Poly.)          & 53.7 & 53.0 & 64.1 & 63.0 & 47.6 & 47.2 & 52.5 & 51.1 \\
SVM (RBF)            & 69.6 & 67.9 & 78.2 & 75.9 & 80.9 & 77.4 & 70.5 & 68.2 \\
SVM (Lin.)           & 50.2 & 48.8 & 57.7 & 56.4 & 47.0 & 45.9 & 48.2 & 47.3 \\
\hline
\end{tabular}
}
    \label{tab:balanced_accuracies}
\end{table}
\begin{table}
\centering
\caption{Aggregated test set confusion matrices across the 5-fold cross validation of the
best performing models. Values are reported rounded to two decimal places.}
\vspace{-2mm}
\setlength{\tabcolsep}{1pt}
\begin{minipage}{0.49\textwidth}
\centering
\caption*{\small (a) Full Annotation -- ResNet18}
\begin{tabular}{lcccc}
\toprule
 & Neutral & Curbing & Overdrive & Edge \\
\midrule
Neutral    & 2555 & 117 & 82  & 18  \\
Curbing    & 159  & 908 & 77  & 88  \\
Overdrive  & 92   & 55  & 803 & 131 \\
Edge       & 24   & 68  & 146 & 1345 \\
\bottomrule
\end{tabular}
\end{minipage}
\hfill
\begin{minipage}{0.49\textwidth}
\centering
\caption*{\small(b) Full Annotation -- ResNet18}
\begin{tabular}{lcccc}
\toprule
 & Neutral & Curbing & Overdrive & Edge \\
\midrule
Neutral    & 0.92  & 0.04 & 0.03  & 0.01 \\
Curbing    & 0.13  & 0.74  & 0.06 & 0.07  \\
Overdrive  & 0.09 & 0.05 & 0.74  & 0.12   \\
Edge       & 0.02 & 0.04 & 0.09 & 0.85   \\
\bottomrule
\end{tabular}
\end{minipage}
\vfill
\begin{minipage}{0.49\textwidth}
\centering
\caption*{\small(c) Nominal Modes – Pret. ResNet34}
\begin{tabular}{lcccc}
\toprule
 & Neutral & Curbing & Overdrive & Edge \\
\midrule
Neutral    & 2514 & 35  & 32  & 1   \\
Curbing    & 49   & 1191 & 20  & 39  \\
Overdrive  & 12   & 13  & 1184 & 24  \\
Edge       & 1    & 17  & 39  & 1470 \\
\bottomrule
\end{tabular}
\end{minipage}
\hfill
\begin{minipage}{0.49\textwidth}
\centering
\caption*{\small(d) Nominal Modes -- Pret. ResNet34}
\begin{tabular}{lcccc}
\toprule
 & Neutral & Curbing & Overdrive & Edge \\
\midrule
Neutral    & 0.97   & 0.014 & 0.01 & 0.00 \\
Curbing    & 0.04  & 0.92  & 0.02 & 0.03    \\
Overdrive  & 0.01 & 0.01 & 0.96  & 0.02   \\
Edge       & 0.00 & 0.03 & 0.01 & 0.96   \\
\bottomrule
\end{tabular}
\end{minipage}
\label{fig:confusion_matrix}
\end{table}

\begin{figure}[!h]
    \centering
    \begin{subfigure}{0.9\linewidth}
 \includegraphics[width=\linewidth]{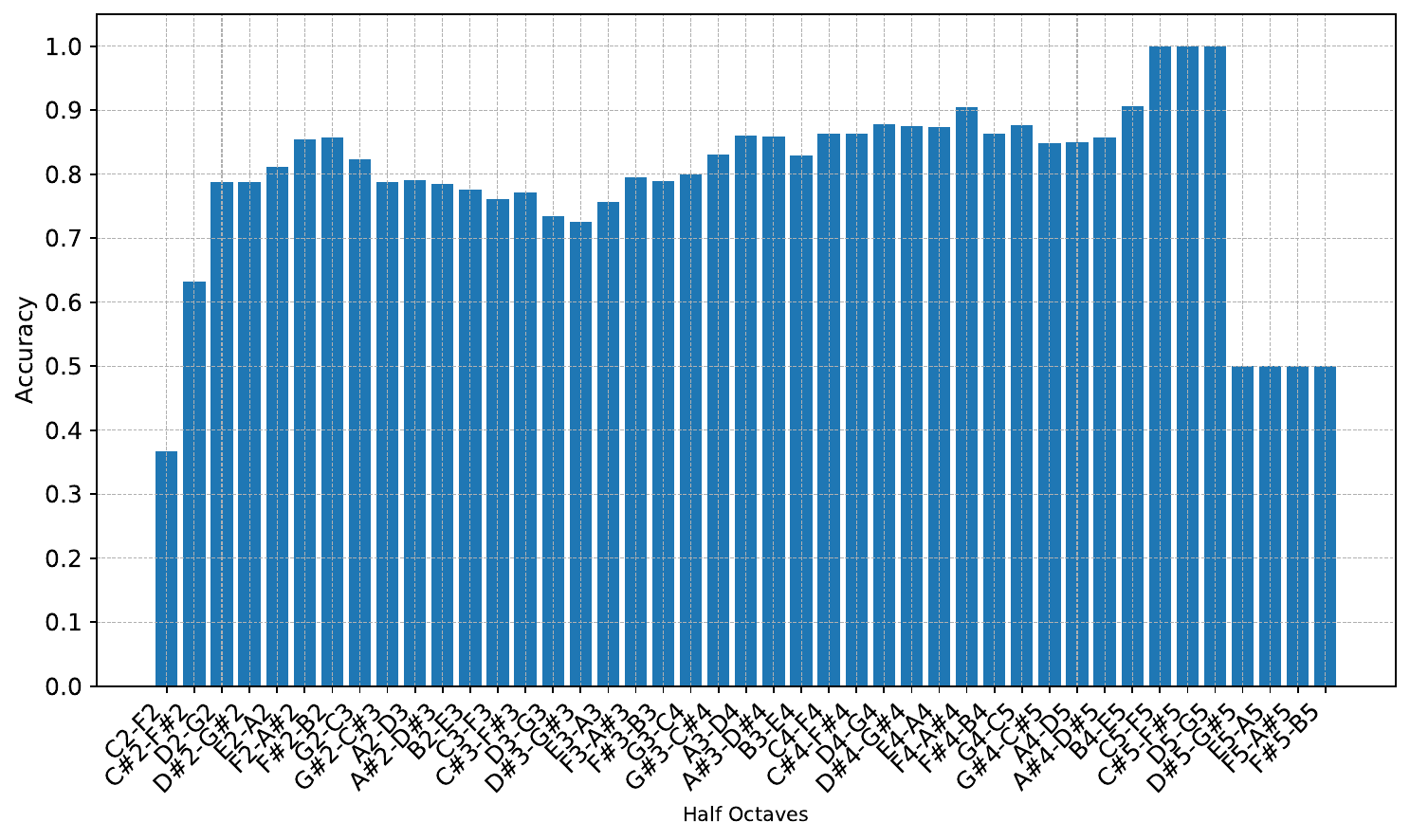} 
        \caption{Using Annotated Mode}
\label{fig:ResNet18_half_octaves}
    \end{subfigure}\hfill
       \begin{subfigure}{0.9\linewidth}
 \includegraphics[width=\linewidth]{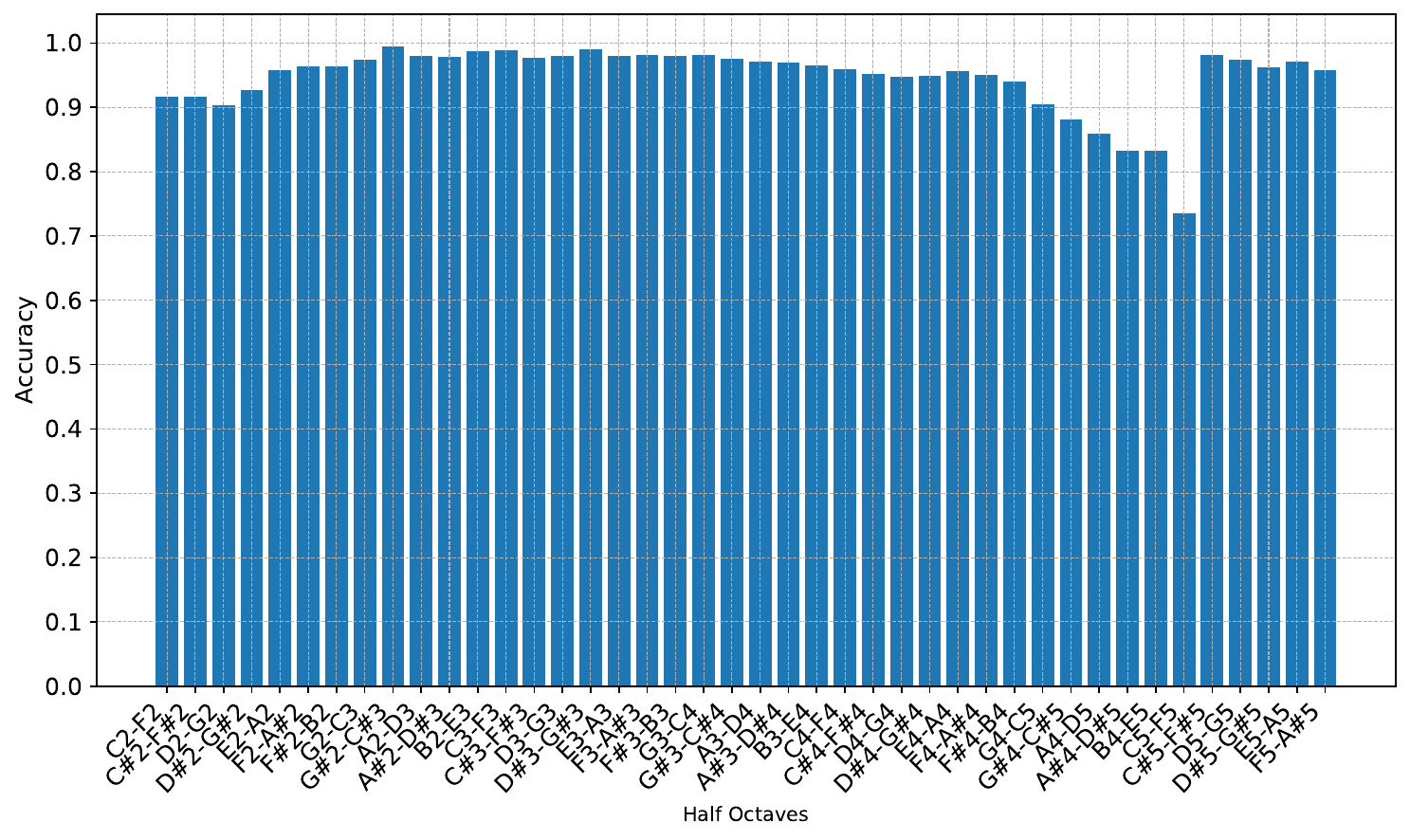}
        \caption{Using Nominal Mode}
\label{fig:ResNet34_half_octaves}
    \end{subfigure}
    \caption{Balanced accuracy across half-octaves on the test set for the best iterations of the 5-fold cross validation when (a) using the full annotation and (b) using the nominal vocal modes as ground truth. The very high accuracy for the top end of the range in (b) is largely due to a lack of data. Correspondingly, the drop to 50\,\% balanced accuracy in (a) is due to a lack of data in this particular iteration, where only for two vocal modes were present for the highest range.    
    Qualitative identical behavior was observed for all classifiers.}
\label{fig:balanced_accuracy_test_half_octaves}
\end{figure}

In the table, results for all four considered annotations/subsets - full annotation, nominal ground truth, strong consensus and weak consensus - are given.
For all these cases, distinct accuracies are reported for the entire respective subset as well as the corresponding phones-only subsets.
The ResNets outperform all other classifiers in every case with XGBoost being the best non-neural classifier.
A significant difference in accuracy is observed between the classification using the annotation (Best accuracy: 81.3\,\%) and the classification based on the nominal vocal modes (Best accuracy: 95.3\,\%).
Aggregated confusion matrices of the test set performance across the 5-fold cross validation are given for the ResNet18, using the full annotation, and the pretrained ResNet34, using the nominal modes as ground truth, in Tab.\ref{fig:confusion_matrix}. 
 Fig. \ref{fig:balanced_accuracy_test_half_octaves} (Appendix) shows exemplary balanced accuracies
across pitch ranges - grouped in half octaves - for the ResNet18 in Fig. \ref{fig:ResNet18_half_octaves}, and for the ResNet34 in Fig. \ref{fig:ResNet34_half_octaves}. 
These results used only the respective best iteration of the cross validation. The ResNet18 was evaluated using the full annotation, whereas the ResNet34 was evaluated using the nominal modes as ground truth.
For the ResNet18, using the full annotation, accuracies varied between 70\,\% and 90\,\% for the most part; they only decayed considerably for the lowest pitch range. 
The apparent loss in performance, starting at D\#5, is an artifact due to low amounts of data in this range and unfortunate sampling in this particular cross validation iteration. Here, samples existed only for two vocal modes, and as such balanced accuracy can at best achieve a value of 50\,\%, as the per class accuracy is assumed zero, if no samples exist. In general, given sufficient data, balanced accuracy was nearly or exactly 100\,\%, as the highest range is especially dominated by Neutral samples.
For the ResNet34, using the nominal vocal modes as ground truth, performance is varying only slightly across pitches, with a minimum of slightly above 70\,\% balanced accuracy for the C5--F5 half octave.

\section{Discussion}
\label{sec:discussion}
This work presented a newly collected dataset for automatic vocal mode classification together with several baseline classification results.\\
\textbf{Dataset}
While the collected dataset should be a reasonable dataset for the general task of automatic vocal mode classification due to the variety in notes and recording devices, additional singers (including amateurs), realistic musical content and singing techniques should be recorded to further address the problem of vocal mode classification.
Additional singers help algorithms to generalize better to unseen ones. Realistic musical content, unlike major arpeggios as for this dataset, would be desirable. This means that the singers actually sing parts of real musical pieces. That way, a more nuanced coverage of the singers' voice and the way the vocal modes can be varied could be captured. However, actually performing such recording sessions is challenging, as singers and musical pieces have to match while maintaining balance between the vocal modes. 
Regarding singing techniques, the voice can be altered in its sound in many ways. Adding more or less air, making it brighter or darker sounding, and using more or less vibrato can drastically change the sound of a voice without changing the vocal mode. 
Such variations are not captured by the dataset but should be so in future work.
Currently, a major obstacle in fully solving automatic vocal mode classification satisfactorily is obtaining a high quality ground truth. Some details of the annotation of this dataset are later discussed.
As the Complete Vocal Institute itself showed \cite{Brixen14}, on average, even very experienced CVT students/teachers do not exceed an accuracy of 80\,\% in acoustic vocal mode classification. However,  \cite{Brixen14} remains unclear about how their ground truth was obtained. By the description, it appears as if the nominal vocal mode served as ground truth. However, in general the nominal vocal mode can be different from the actual vocal mode, as, e.g., no proper hold was established for Curbing potentially resulting in Overdrive-like samples.
A jury of highly experienced CVT-informed singers is required, out of which those with the highest test-retest ratings should be used to determine a definite ground truth similar to the procedure of \cite{Saldías24}. However, considerable funding is required to achieve this.\\
\textbf{Classifier Performance}
While the achieved balanced accuracies of slightly above 80\,\% for the annotation, or up to about 95\,\% when using the nominal modes, is a good to very good performance, one cannot expect these accuracies to generalize to arbitrary, unseen samples due to the mentioned limitations of the dataset in terms of vocal traits.
When using the full annotation, the observed decline in balanced accuracy for the lowest notes of the dataset, observed for all classifiers investigated, can be explained by the following reasons: 
The vocal modes according to CVT can be sung  in a wide note range, with certain limits depending on vocal mode. Nonetheless, certain ranges are more common than others, resulting in the subjects being more accustomed to producing the vocal modes in certain ranges. Thus, not many, e.g., Curbing samples were recorded for the lowest note range. Furthermore is it uncommon to hear all vocal modes in very low vocal ranges. Because the sound properties typically change somewhat with changing pitch, it might be difficult to distinguish the vocal modes for very deep vocals without special training or at all. Further research is required to investigate this point.\\
\textbf{Performance for the Different Subsets}
The substantial decrease in balanced accuracy when using the annotation compared to using the nominal vocal modes might be evidence that the singers are better "judges" than the annotators/listeners. The singers had the advantage of being able to "carry" the vocal modes from the high range down to the lower range, as they were using major arpeggios in one go across their entire vocal range.
This hypothesis is supported by the observed improvement of the classification accuracy, when moving from the full annotation to the strong consensus subset, where the classifiers achieved only slightly worse balanced accuracies, with the best classifier achieving a balanced accuracy of 90.9\,\% compared to 94.9\,\% for the nominal vocal mode problem and 80.8\,\% when using the full annotation.
Also in line with this, is the considerably decreased classifier performance on the weak consensus subset. As the additional samples with slight disagreement between the annotators could introduce conflicting annotations - highly similar acoustic features get different ratings -, it is no surprise to see a decrease in classifier performance.
However, the comparison is somewhat impeded by the reduced number of samples for the strong consensus subset.
The hypothesis, based on the observed classifier accuracies, then is, that the acoustic features distinguishing the vocal modes are consistently present in the recorded samples across most of the vocal ranges of the singers. This is suggested by the high balanced accuracy for the nominal vocal modes.  However, these features are in general potentially either difficult to hear or the modes are rarely heard for certain ranges, leading to confusion by the annotators.
If this is the case, then only additional annotators can ameliorate the situation.
As little to no decrease in balanced accuracy was observed for the respective phone subsets, the classifiers should be somewhat robust to the used recording device and slight alterations of the room acoustics. This is useful for future smart-device applications. 
Prior tests, which used simpler convolutional neural networks and the nominal modes as ground truth, showed a considerable drop in balanced accuracy - more than 25\,\% for the MotorolaOne samples - for the phone subset, if the models were trained solely on the data recorded with the Behringer microphone.\\
\textbf{Comparison to Related Work}
The two published previous attempts at automatic vocal mode classification showed subpar classification performance. For \cite{Brixen13}, the classifier model - a two step decision tree based on vocal harmonics -  was too simple. Additionally, the amount of data recorded was limited, with a single scale per vowel and four singers in total.
For \cite{Sol23}, which achieved 68.6\,\% balanced test accuracy for vocal mode classification using several features and the XGBoost algorithm, the amount of data, a crucial point in virtually all machine learning applications, was likely insufficient to achieve high accuracy. Each singer supplied only three different pitches. Additionally, recording was done by the subjects themselves, likely increasing classification complexity due to different recording conditions. 
The best comparison to \cite{Sol23} is our result when using the nominal modes as ground truth. For this, we achieved a balanced accuracy of 95.3\,\% using a pretrained ResNet34, aggregated across the 5-fold cross validation identically to \cite{Sol23}. This represents a gain of more than 25\,\% in balanced accuracy. Even using the exact same classifier that was used in \cite{Sol23}, namely XGBoost, the gain in balanced accuracy is larger than 20\,\%, as we achieved a balanced accuracy of 92.8\,\% with XGBoost.
However, due to the different datasets and slightly different ways the ground truth was obtained, a direct comparison is at least slightly unfair. It would be interesting to assess our models' performances on the data recorded for \cite{Sol23}.

\subsection{Quality of the Annotation}
\label{ssec:quality_annotation}
\begin{figure}
    \centering
    \includegraphics[width=\linewidth]{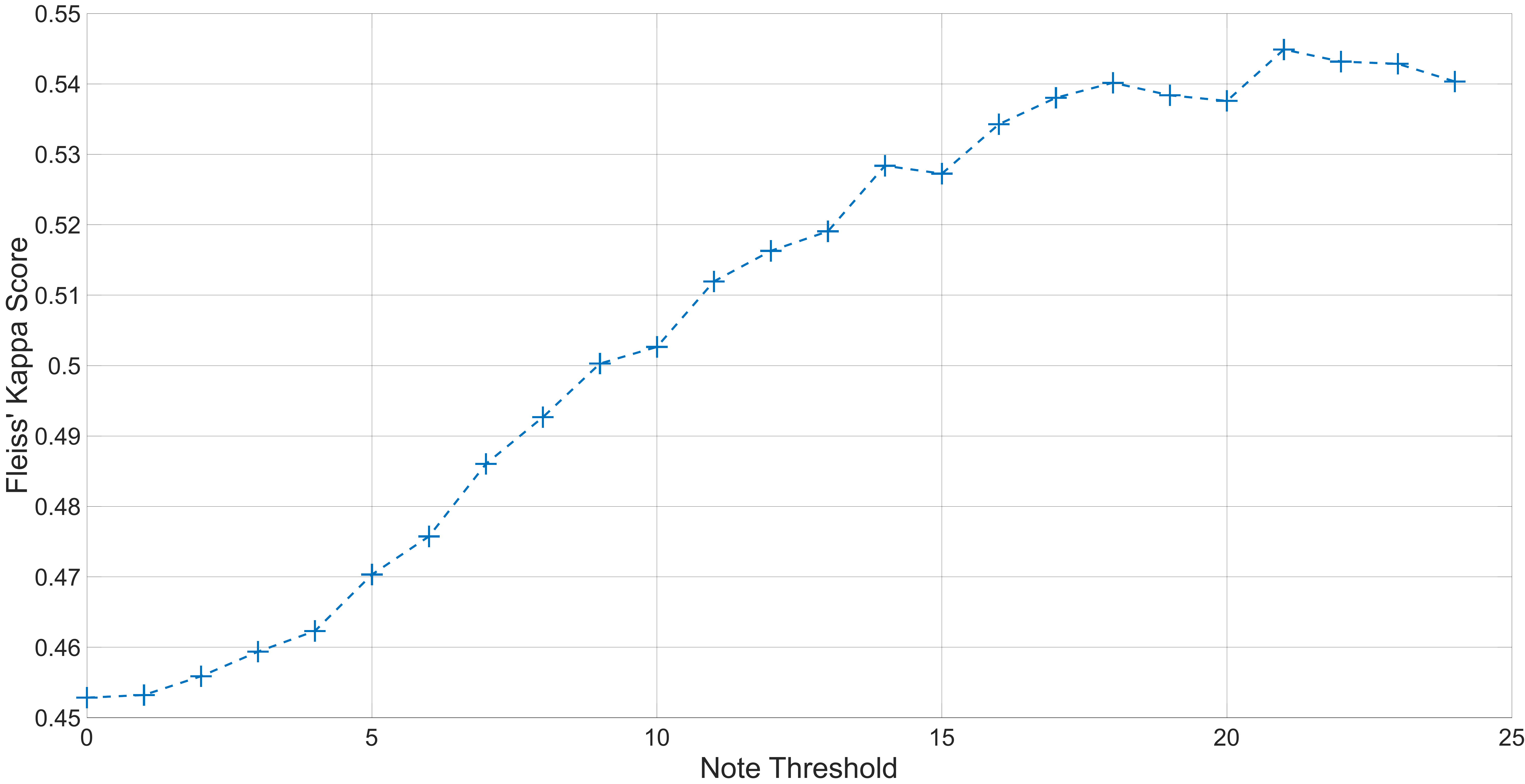}
    \caption{Fleiss' kappa score across cut-off note threshold. The computation of the Fleiss' kappa score given a note threshold of K ignores the lowest K notes of each singer, irrespective of the vocal mode. }
    \label{fig:fleiss_across_threshold}
\end{figure}

A Fleiss' kappa score of 0.45 indicates a substantial amount of disagreement between the annotators. Part of this disagreement stems from vocal modes being sung in lower registers. Here, both annotators and singers are possibly less experienced, as Overdrive and Edge, for instance, are most prominently heard in high notes, usually as part of a climax in a musical piece.
This claim is substantiated by Fig. \ref{fig:fleiss_across_threshold} (Appendix), which depicts the Fleiss' kappa score for a subset of the dataset, namely those resulting from ignoring the lowest K notes - and all corresponding samples - measured in semitones, of each singer (irrespective of the vocal mode), where K ranges from 0 semitones to 24 semitones.
By increasingly ignoring lower notes, the Fleiss' kappa score rises to around 0.54, about 0.09 higher than for the entire annotation.
A tiny part of the Fleiss' kappa score could additionally be explained by human oversight in writing down the annotation. Excel sheets with four columns, one per vocal mode, were used, where an 'x' was used to mark the respective annotation. It is possible that in isolated instances, by accident, the wrong column was marked. 
Another reason for disagreements might be fatigue, as annotating large numbers of samples can be exhausting, such that later samples may have been rated with less scrutiny. Because the annotators created their annotation at home, the amount of samples rated in one session was not controlled.
In the future, splitting annotation sessions into calibration - annotation - rest cycles could presumably improve annotation consistency. Calibration and rest cycles should allow to maximize intra-rater agreement and as such annotation consistency, i.e., giving the same rating for identical or virtually identical features. 
Unfortunately, neither \cite{Sol23} nor \cite{Brixen14} reported a Fleiss' kappa score, making it difficult to draw comparisons to these publications.
It can only be assumed to be similar based on their reported agreement between the nominal and the annotated modes, which are only slightly lower than the one reported in this work.
Especially for lower pitches, additional annotators are of great interest. It is hoped that others, in particular the Complete Vocal Institute, are making attempts to contribute to our annotation, such that the quality of the annotation can be improved through a larger number of additional annotators. 
Crowd sourcing, both for recording samples and annotation, could be feasible, however, it was attempted to find additional singers through the CVT network to no avail.

\subsection{Limitations}
Due to a lack of interested professional singers, the main author also participated as a vocalist. As the recordings were supposed to be performed at the main authors' institute, only local professionals could realistically be invited. 

While the main author did participate in the study by providing voice recordings in the same way the other subjects did, with s1 guiding this recording session, the main author did not influence the annotation itself. The main author did not provide an annotation. The individual annotations were created entirely during the private time of the three annotators without the presence of the main author.
All annotators were explicitly encouraged to judge the anonymized samples according to their own personal judgment.

Therefore it is not assumed that the participation of the main author in the recording session impacted the annotation of the dataset.

\section{Conclusion}
This work presents a novel dataset for automatic vocal mode classification as well as baseline classification results. Four subjects (two male, two females) three of whom professional singers with extensive CVT-experience, were recorded resulting in 3,752 unique samples, each covering a single sustained note, covering the entire vocal range of the singers. These samples are made up of 1,605 nominal Neutral, 642 nominal Curbing, 664 nominal Overdrive and 841 nominal Edge examples. In total, four microphones, including two smartphones, were used to record the subjects, resulting in more than 13,000 samples in total.  Three annotators provided a ground truth annotation.
 The best balanced test set accuracy of 81.3\,\% was achieved by a ResNet18. The best balanced test set accuracy rose to 95.3\,\%, achieved by the ResNet34, if the nominal vocal modes were used as ground truth.

\section*{Acknowledgement}
The authors would like to thank all invited subjects for contributing to our dataset. Furthermore we would like to thank Samantha Nicole Carbajal P{\'e}rez for the initial data segmentation of the Behringer recordings. The authors would also like to thank all anonymous reviewers for their helpful comments.

\section*{Conflict of Interest}
The authors declare no conflict of interest. This research
did not receive any specific grant from funding agencies in
the public, commercial, or not-for-profit sectors. Due to a lack of interested professional singers, the main author also participated as a vocalist.

\bibliographystyle{splncs04}
\bibliography{refs}
%




\newpage
\section*{Appendix}

\FloatBarrier

\end{document}